\documentclass[aps,prd,twocolumn,nofootinbib,showpacs]{revtex4-1}

\usepackage{eqnarray,graphicx,epsfig}
\usepackage[colorlinks=true,linkcolor=blue,citecolor=blue, urlcolor=blue]{hyperref}

\begin{document}

\title{Properties of the Free Energy Density Using the Principle of Maximum Conformality}

\author{Shi Bu}
\author{Xing-Gang Wu} \email{wuxg@cqu.edu.cn}
\author{Jian-Ming Shen}
\author{Jun Zeng}

\address{Department of Physics, Chongqing University, Chongqing 401331, P.R. China}

\begin{abstract}
We present a detailed study on the properties of the free energy density at the high temperature by applying the principle of maximum conformality (PMC) scale-setting method within the effective field theory. The PMC utilizes the renormalization group equation recursively to identify the occurrence and pattern of the non-conformal $\{\beta_i\}$-terms, and determines the optimal renormalization scale at each order. Our analysis shows that a more accurate free energy density up to $g_s^5$-order level without renormalization scale dependence can be achieved by applying the PMC. We also observe that by using a smaller factorization scale around the effective parameter $m_E$, the PMC prediction shall be consistent with the Lattice QCD prediction derived at the low temperature.

\pacs{12.38.Aw, 12.38.Bx, 12.38.Cy}

\end{abstract}

\maketitle

\section{Introduction}
\label{sec:introduction}

At extremely high temperature, the hadronic matter are assumed to occur a phase transition to the quark-gluon plasma (QGP). The QGP might come from the early universe up to a few milliseconds after the Big Bang or from the heavy ion collisions at the Relativistic Heavy Ion Collider (RHIC) and the Large Hadronic Collider (LHC), and etc. This system therefore behaves more like a collection of free quarks and gluons rather than a collection of their bound states~\cite{Kajantie:2000iz}.

The static equilibrium properties of the QGP at the temperature $T$ are governed by the free energy density~\cite{Braaten:1995jr}
\begin{equation}\label{eq:Fdefinition}
F=-\frac{T}{V}\ln {\cal Z}_{\rm QCD},
\end{equation}
where $V$ is the space volume and the partition function ${\cal Z}_{\rm QCD}$ is a functional integral over quark and gluon fields on a $4$-dimensional Euclidean space-time, with the Euclidean time taking its values on a circle with circumference $1/T$. In the limit when the quarks are massless, the free energy density is a function of $T$ and the strong coupling constant.

During the past decades, the free energy density, or equivalently the negative pressure, of the QGP has been calculated by using the lattice gauge theory~\cite{Boyd:1996bx, Okamoto:1999hi, Bernard:2004je, Aoki:2005vt, Bernard:2006nj, Cheng:2007jq, Endrodi:2007tq, DiRenzo:2008en, Hietanen:2008tv, Bazavov:2009zn, Borsanyi:2010cj, Borsanyi:2011zm, Borsanyi:2012cr, Borsanyi:2012ve, Borsanyi:2012uq, Bazavov:2012jq, Bazavov:2012vg, Philipsen:2012nu, Borsanyi:2013hza, Bazavov:2013dta, Gursoy:2016ebw} or the perturbative QCD (pQCD) theory~\cite{Shuryak:1977ut, Kapusta:1979fh, Toimela:1982hv, Arnold:1994ps, Arnold:1994eb, Zhai:1995ac, Braaten:1995jr, Braaten:1995ju, Kajantie:2002wa}. In the present paper, we shall focus on the circumstance that the QGP has a high temperature $T$ ($T$ is considered as a measure of the average energy of the constituents), indicating the quarks and gluons are of high energy and the strong couplings among them are small due to asymptotic freedom. Within this temperature region, the pQCD theory is a feasible tool to study the free energy density. During the calculation, we shall resum specific diagrams such as the ``ring diagrams"~\cite{Zhai:1995ac, Braaten:1995jr}, and it is helpful to expand the perturbative series by the coupling constants $g_s$ rather than $\alpha_s$. The free energy density at the high tempeture has been calculated up to ${\cal O}(g_s^2)$~\cite{Shuryak:1977ut}, ${\cal O}(g_s^3)$~\cite{Kapusta:1979fh}, ${\cal O}\left(g_s^4 \ln(1/{g_s})\right)$~\cite{Toimela:1982hv}, ${\cal O}(g_s^4)$~\cite{Arnold:1994ps, Arnold:1994eb}, ${\cal O}(g_s^5)$~\cite{Zhai:1995ac, Braaten:1995jr, Braaten:1995ju}, and part of ${\cal O}(g_s^6)$~\cite{Kajantie:2002wa}, respectively. The ${\cal O}(g_s^0)$ term is the free energy density of the ideal gas. The ${\cal O}(g_s^2)$ and higher-order terms contain the corrections from the interactions among the basic particles, the screening effects from the plasma, and etc. There are new nonperturbative effects entangled with the infrared divergence emerge at the ${\cal O}(g_s^6)$-order~\cite{Linde:1978px, Linde:1980ts, Gross:1980br}, and at present, only the specific terms of the form ${\cal O}\left(g_s^6\ln{g_s}\right)$ have been achieved.

For a high-order pQCD prediction, one has to choose a renormalization scheme and a renormalization scale $\mu_r$ to finish the renormalization. The scale $\mu_r$ is usually taken as the typical momentum flow of the process or the one to eliminate the large logs such that to make the pQCD series relatively steady over the scale changes. For the present case, one usually sets $\mu_r=2\pi T$, which corresponds to the energy of the first non-vanishing Matsubara mode~\cite{Matsubara:1955ws}. However, such a simple choice of ``guessed" scale leads to the miss-matching of the perturbative coefficients to the strong coupling constant, resulting in the well-known scheme-and-scale ambiguities persist at any fixed order~\cite{Celmaster:1979km, Abbott:1980hwa, Buras:1979yt, Grunberg:1980ja, Stevenson:1981vj, Brodsky:1982gc}. By using the ``guessed" scale, there are some other defects~\cite{Wu:2013ei, Wu:2014iba}, especially, I) The predictions for a guessed scale are incorrect for Quantum Electrodynamics, whose renormalization scale can be unambiguously set by the Gell-Mann-Low procedure~\cite{GellMann:1954fq}; II) The perturbative series is factorially divergent at large order -- the renormalon problem~\cite{Beneke:1998ui, Gardi:2001wg}; III) It is often argued that such scale uncertainties can be suppressed by including enough high-order terms, which however shall be diluted by the divergent renormalon terms; IV) If a poor pQCD convergence is observed for an observable, one can not decide whether it is the intrinsic property of pQCD series or is caused by improper choice of scale.

Many attempts have been tried to improve the prediction on the free energy density~\cite{Karsch:1997gj, Chiku:1998kd, Andersen:1999fw, Andersen:1999va, Andersen:2000yj, Kastening:1997rg, Hatsuda:1997wf, Cvetic:2002ju, Cvetic:2004xq, Hatsuda:1994pi, Fukushima:2003fm, Fukushima:2003fw, Ratti:2005jh, Mukherjee:2006hq, Bhattacharyya:2010wp, Bhattacharyya:2010ef, Bluhm:2007cp, Bannur:2006ww, Bannur:2007tk, Gardim:2009mt, Schaefer:2009st, Schaefer:2009ui, Skokov:2010uh}. In the paper, we will apply the principle of maximum conformality (PMC)~\cite{Brodsky:2011ta, Brodsky:2012rj, Mojaza:2012mf, Brodsky:2013vpa} to the free energy density up to ${\cal O}(g_s^5)$ with the goal of eliminating the renormalization scale ambiguity and achieving an accurate pQCD prediction which is independent of theoretical conventions. Because the running behavior of the coupling constant is controlled by the renormalization group equation (RGE) or the $\beta$-function, the PMC suggests to use the knowledge of the $\{\beta_i\}$-terms from the known pQCD series to determine the optimal scale of a particular process. A recent review on this point can be found in Ref.\cite{Wu:2018cmb}. If one fixes the renormalization scale of the pQCD series using the PMC, all the non-conformal $\{\beta_i\}$-terms in the perturbative series shall be resummed into the running coupling, one thus obtains a unique, scale-fixed, and scheme-independent prediction at any fixed order. Many PMC applications have been done in the literature, cf. the review~\cite{Wu:2015rga}, all of those examples show that due to the rapid convergence of conformal pQCD series, the residual uncertainties are highly suppressed, even for low-order predictions.

There are several typical momentum flows for the free energy density up to ${\cal O}(g_s^6)$, e.g. $T$, ${g_s}T$, $g_s^2 T$~\cite{Ginsparg:1980ef, Appelquist:1981vg, Nadkarni:1982kb}. The QCD effective field theory (EFT) provides a systematic way to unravel the contributions under different energy scales. The QCD EFT is a three-dimensional one in which all the quarks and non-static bosons have been integrated out of the theory such that it reduces to purely static bosonic modes~\cite{Ginsparg:1980ef, Appelquist:1981vg, Nadkarni:1982kb, Braaten:1995jr, Braaten:1995ju}. The EFT factorizes the free energy density of the hot QCD into the perturbative coefficients and the non-perturbative parts via proper matching. The PMC can be applied separately to set the renormalization scale of the free energy density within different scale regions.

The remaining parts of the paper are organized as follows. In Sec.~\ref{sec:calculation}, we present the calculation technology for achieving the PMC prediction on the free energy density. Numerical results and discussions are presented in Sec.~\ref{sec:results}. Sec.~\ref{sec:summary} is reserved for a summary.

\section{Calculation technology}
\label{sec:calculation}

Using the EFT, the free energy density can be decomposed into various parts which are characterized by typical scales as $T$, $g_s T$ and $g_s^2 T$, and they are labeled as $F_E$, $F_M$, and $F_G$, respectively. Here the hard part $F_E$ can be treated as a power series in $\alpha_s=g_s^2/4\pi$, the softer part $F_M$ is a power series in $g_s$ which begins at the $g_s^3$-order, and the softest part $F_G$ is a power series in $g_s$ which begins at the $g_s^6$-order. At present, the complete $g_s^6$-order terms are not known, so we shall concentrate our attention on the free energy density up to $g_s^5$-order.

Up to $g_s^5$-order, the free energy density $F$ can be formulated as~\cite{Braaten:1995jr, Braaten:1995ju},
\begin{equation}\label{eq:FreeEnergy}
F=F_E(\Lambda_E)+F_M(\Lambda_E),
\end{equation}
where $\Lambda_E$ is the factorization scale. The hard part $F_E(\Lambda_E)$ can be expressed as
\begin{equation}\label{eq:tilFE0}
F_E(\Lambda_E)=F_{\rm ideal}+\frac{8\pi^2}{3}{T^4}{\widetilde F}_E(\Lambda_E),
\end{equation}
where $F_{\rm ideal}$ stands for the contribution of the ideal quark-gluon gas
\begin{equation}\label{eq:Fideal}
F_{\rm ideal}=-\frac{8\pi^2}{45}{T^4}\left(1+\frac{21}{32}{n_f}\right),
\end{equation}
and ${\widetilde F_E}(\Lambda_E)$ represents the ``canonic'' QCD part,
\begin{eqnarray}
{\widetilde F}_E(\Lambda_E) = F'_E - \left[ 144\left(1+\frac{1}{6}{n_f}\right)\ln\frac{\Lambda_E}{2\pi T} \right] a_s^2(\mu_r),
\label{eq:tilFE}
\end{eqnarray}
where $\mu_r$ is the (arbitrary) renormalization scale. The remaining part $F'_E$ is perturbatively calculable, which can be expressed as
\begin{eqnarray}\label{perturbativeFE}
F'_E = r_{1,0}^E{a_s}(\mu_r)+\left(r_{2,0}^E + r_{2,1}^E{\beta_0}\right)a_s^2(\mu_r),
\end{eqnarray}
where $a_s=\alpha_s/4\pi$, $\beta_0= 11-2/3{n_f}$ with $n_f$ being the active flavor numbers emerged in the $\alpha_s$-renormalization. As required by the PMC, we have transformed those $n_f$-terms into the $\{\beta_i\}$-series. The conformal coefficients $r_{i,j(=0)}^E$ and the nonconformal ones $r_{i,j(\neq0)}^E$ under the $\rm{\overline{MS}}$-scheme read
\begin{eqnarray}
r_{1,0}^E &=& 1+\frac{5}{12}{n_f}, \\
r_{2,0}^E &=& -214.54-29.15\left(1+\frac{5}{12}{n_f}\right), \\
r_{2,1}^E &=& \left(-1.59+2\ln\frac{\mu_r}{2\pi T}\right)\left(1+\frac{5}{12}{n_f}\right).
\end{eqnarray}
Here the $n_f$-terms in those coefficients $r_{i,j}$ are free quark numbers in QGP, which are irrelevant to the running of the coupling constant and should be kept as conformal coefficients when applying the PMC~\cite{Brodsky:2011ta, Brodsky:2012rj, Mojaza:2012mf, Brodsky:2013vpa}.

After applying the PMC, the pQCD series of $F'_E$, e.g. Eq.(\ref{perturbativeFE}), can be improved as the following scheme-independent conformal series,
\begin{eqnarray}\label{eq:FEpmc}
F'_E = r_{1,0}^E {a_s}(Q^{e}_1) + r_{2,0}^E a_s^2(Q^{e}_1),
\end{eqnarray}
where $\ln (Q^e_1)^2/\mu_r^2 = -r_{2,1}^E/r_{1,0}^E$. We have set the NLO PMC scale ${Q^e_2}={Q^e_1}$ to ensure the scheme independence, whose exact value can be determined by using the NNLO terms which are not available at the present.

The softer part ${F_M}(\Lambda_E)$ can be expressed by using the EFT parameters, $m_E^2$ and $g_E^2$, as~\cite{Braaten:1995jr, Braaten:1995ju}
\begin{eqnarray}
{F_M}(\Lambda_E) &=& -\frac{2T}{3\pi}m_E^3 \bigg[ 1- \left(0.256+\frac{9}{2}\ln\frac{\Lambda_E}{m_E}\right)\frac{g_E^2}{2\pi{m_E}} \nonumber \\
&& - 27.6\left(\frac{g_E^2}{2\pi{m_E}}\right)^2 \bigg] .  \label{FMlam}
\end{eqnarray}

In deriving $F_M(\Lambda_E)$, the RGE-involved fermion-loop contributions have been incorporated into the EFT parameters $m_E^2$, $g_E^2$, etc.~\cite{Ginsparg:1980ef, Appelquist:1981vg, Nadkarni:1982kb, Braaten:1995jr, Braaten:1995ju}, therefore it is better to apply the PMC directly to those parameters such that to get more accurate prediction on those EFT parameters and to avoid the double counting problem~\footnote{The PMC resums the $\{\beta_i\}$-terms into the running coupling and the pQCD series of the free energy density also includes partially resummation effects~\cite{Arnold:1994ps, Arnold:1994eb, Zhai:1995ac}. A straightforward application of PMC to the pQCD series might contain a mixing of inequivalent resummations, leading to double counting problem.}.

To be consistent with the known $g_s^5$-order prediction for the free energy density, we need to know $m_E^2$ up to the next-to-leading order (NLO) level and $g_E^2$ to the leading-order (LO) level. The $m_E^2$ up to NLO level~\cite{Braaten:1995jr, Braaten:1995ju} can be written as
\begin{eqnarray}
m_E^2 = 16{\pi^2}{T^2} \left[ r_{1,0}^m a_s(\mu_r) + \left(r_{2,0}^m+r_{2,1}^m{\beta_0}\right)a_s^2(\mu_r)\right],
\label{eq:mE}
\end{eqnarray}
where $r_{i,j}^m$ under the $\rm{\overline{MS}}$-scheme read
\begin{eqnarray}
r_{1,0}^m &=& 1+\frac{1}{6}{n_f}, \\
r_{2,0}^m &=& 8-22.50\left(1+\frac{1}{6}{n_f}\right), \\
r_{2,1}^m &=& \left(1.54+2\ln\frac{\mu_r}{2\pi T}\right)\left(1+\frac{1}{6}{n_f}\right).
\end{eqnarray}
Here the $n_f$-terms in those coefficients $r_{i,j}$ are again free quark numbers in QGP. After applying the PMC scale-setting, we obtain
\begin{eqnarray}\label{eq:mEpmc}
m_E^2=16{\pi^2}{T^2}\left[r_{1,0}^m{a_s}(Q_1^m)+r_{2,0}^m a_s^2(Q_1^m)\right],
\end{eqnarray}
where $\ln(Q_1^m)^2/\mu_r^2 = -r_{2,1}^m/r_{1,0}^m$.

By using the LO $g_E^2$ alone, we cannot determine its renormalization scale, and to ensure the scheme independence of $F_M(\Lambda_E)$ at the $g_s^5$-order level, we directly set its value as $Q_1^g=Q_1^m$.

By using the known NLO-terms for $g_E^2$, we can determine its optimal scale by applying the PMC via the same way. For example by using the computed $g_s^6$-order terms from Ref.\cite{Kajantie:2002wa}, we obtain $\ln(Q_1^g)^2/\mu_r^2 = -r_{2,1}^g/r_{1,0}^g$ with the coefficients
\begin{eqnarray}
r_{1,0}^g &=& 1, \\
r_{2,0}^g &=& -29.5, \\
r_{2,1}^g &=& 2.54+2\ln\frac{\mu_r}{2\pi T}
\end{eqnarray}
and
\begin{eqnarray}\label{eq:gE}
g_E^2 = 16{\pi^2}{T} \left[ r_{1,0}^g a_s(Q_1^g) + r_{2,0}^g a_s^2(Q_1^g) \right].
\end{eqnarray}

As a summary, our final prediction for the free energy density $F$ with the factorization scale $\Lambda_E=2\pi T$ is
\begin{eqnarray}\label{eq:Fpmc}
F &=& F_{\rm ideal}+\frac{8\pi^2}{3}{T^4}\left[r_{1,0}^E {a_s}(Q_1) + r_{2,0}^E a_s^2(Q_1)\right] \nonumber \\
&& -\frac{2T}{3\pi}m_E^3 \bigg[ 1- \left(0.256+\frac{9}{2}\ln\frac{2\pi T}{m_E}\right)\frac{g_E^2}{2\pi{m_E}} \nonumber \\
&& - 27.6\left(\frac{g_E^2}{2\pi{m_E}}\right)^2 \bigg].
\end{eqnarray}
If choosing the factorization scale $\Lambda_E=m_E$, we obtain
\begin{eqnarray}\label{eq:Fpmc2}
F &=& F_{\rm ideal}+\frac{8\pi^2}{3}{T^4}\left[r_{1,0}^E {a_s}(Q_1) + r_{2,0}^E a_s^2(Q_1)\right] \nonumber \\
&& - \frac{8 \pi^2}{3} T^4 \left[ 144\left(1+\frac{1}{6}{n_f}\right)\ln\frac{m_E}{2\pi T} \right] a_s^2(Q_1) \nonumber\\
&& -\frac{2T}{3\pi}m_E^3 \bigg[1-0.256\frac{g_E^2}{2\pi{m_E}} - 27.6\left(\frac{g_E^2}{2\pi{m_E}}\right)^2 \bigg].
\end{eqnarray}

\section{Numerical results and discussions}
\label{sec:results}

To do the numerical calculation, we adopt the value, $\alpha_{s}( {1.5GeV,{n_f} = 3}) = 0.336_{- 0.008}^{+ 0.012}$~\cite{Bazavov:2014soa}, as a reference point to determine the QCD asymptotic scale $\Lambda_{\overline{\rm MS}}$. By using the two-loop $\alpha_s$-running formulae, we obtain $\Lambda _{\overline {\rm MS} }^{{n_f} = 3} = 0.343_{- 0.012}^{ + 0.018}$ GeV. If not specially stated, we shall adopt $\Lambda_{E}=2\pi T$ as the default value of the factorization scale. In the following, we shall set the temperature $T=10$ GeV as an example to show the basic properties of the free energy density.

\begin{figure}[htb]
\centering
\includegraphics[width=0.5\textwidth]{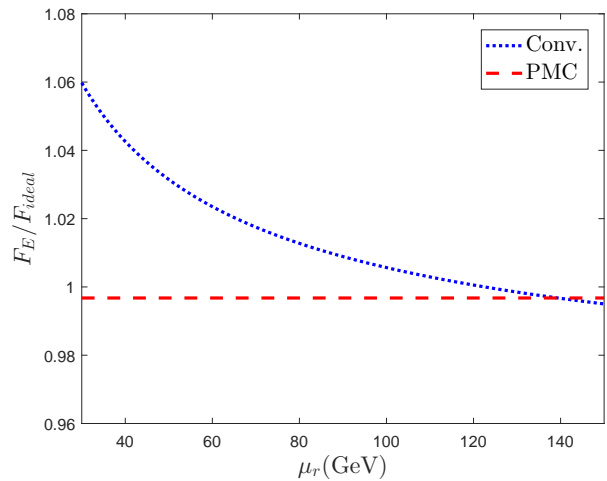}
\caption{The ratio ${F_E}/{F_{\rm ideal}}$ versus the renormalization scale ($\mu_r$) under conventional (Conv.) and PMC scale-settings, respectively. $T=10$ GeV.}
\label{fig:FE}
\end{figure}

\begin{table}[htb]
\begin{center}
\begin{tabular}{c cccccc}
\hline
${F_E}/{F_{\rm ideal}}$~ & ~$\mu_r$ ~ & ~ LO~ & ~NLO~ & ~NNLO~ & ~Total~ \\
\hline
       & ${\pi T}$ & $1$ & $-0.11$ & $0.17$ & $1.06$  \\
$\rm{Conv.}$ & ${2\pi T}$ & $1$ & $-0.10$ & $0.12$ & $1.02$  \\
       & ${4\pi T}$ & $1$ & $-0.09$ & $0.09$ & $1.00$  \\
\hline
$\rm{PMC}$  & $\left[ {\pi T,4\pi T} \right]$ & $1$ & $-0.09$ & $0.08$ & $0.99$ \\
\hline
\end{tabular}
\caption{The ratio ${F_E}/{F_{\rm ideal}}$ under conventional (Conv.) and PMC scale-settings. Three typical renormalization scales, ${\mu_r} = \pi T$, $2\pi T$ and $4\pi T$, are adopted. $T=10$ GeV.}
\label{tab:FE}
\end{center}
\end{table}

Firstly, we discuss the properties of the hard part ($F_E$) of the free energy density, which is characterized by the scale around $T$. We present the renormalization scale dependence of the ratio ${F_E}/{F_{\rm ideal}}$ before and after applying the PMC in Fig.\ref{fig:FE}. We present the numerical results for the ratio $F_E/F_{\rm ideal}$ under several typical choices of renormalization scale, $\mu_r=\pi T$, $2\pi T$ and $4\pi T$, in Table~\ref{tab:FE}. After applying the PMC, $F_E$ is independent to the choice of $\mu_r$, while the NNLO prediction under conventional scale-setting still shows a strong scale dependence. For example, Table~\ref{tab:FE} shows ${F_E}/{F_{\rm ideal}}$ varies by $[-2\%,+4\%]$ for $\mu_r\in[\pi T, 4\pi T]$. It is interesting to find that the typical momentum flow of $F_E$ should be $\simeq 4\pi T$, at which the PMC and conventional scale-settings get almost the same prediction, which is different from the usually considered $2\pi T$ by about two times. This condition is similar to the observation that the preferable choice of the renormalization scale for $gg\to H$ or $H\to gg$ is $m_H/4$~\cite{Wang:2016wgw} and the preferable one for $H\to \gamma\gamma$ is $2m_H$~\cite{Wang:2013akk}, other than the usually considered $m_H$. The typical momentum flow under conventional scale-setting is usually approximated by eliminating the large log-terms of the perturbative series, while the PMC scale-setting provides a reliable way to set the exact value for the typical momentum flow for high-energy process.

Secondly, we consider the properties of the softer part ($F_M$) of the free energy density, which is characterized by a softer scale around $g_s T$.

\begin{figure}[htb]
\centering
\includegraphics[width=0.5\textwidth]{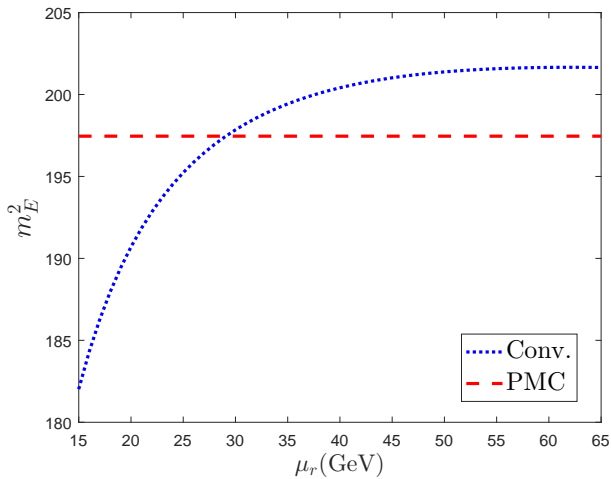}
\caption{The EFT parameter $m_{\rm E}^2$ versus the renormalization scale ($\mu_r$) under conventional and PMC scale-settings, respectively. $T=10$ GeV.}
\label{fig:mE}
\end{figure}

\begin{table}[htb]
\begin{center}
\begin{tabular}{c ccccc}
\hline
$m_{\rm E}^2$~ & ~${\mu_r}$ ~ & ~LO~ & ~NLO~ & ~Total~ \\
\hline
& ${\pi T/2}$ & $271.77$ & $-88.09$ & $183.68$  \\
~\rm{Conv.}~ & ${\pi T}$ & $235.22$ & $-36.85$ & $198.37$  \\
& ${2\pi T}$ & $207.66$ & $-6.00$ & $201.66$  \\
\hline
~PMC~ & $\left[ {\pi T/2,2\pi T} \right]$ & $238.78$ & $-41.32$ & $197.46$  \\
\hline
\end{tabular}
\caption{The results of $m_{\rm E}^2$ under conventional and PMC scale-settings, respectively. Three typical scales, $\mu_r=\pi T/2$, $\pi T$, and $2\pi T$, are adopted. $T=10$ GeV.}
\label{tab:mE}
\end{center}
\end{table}

To show how the scale uncertainty of the EFT parameter such as $m_{\rm E}^2$ changes, we vary the renormalization scale from ${g_s}T\left(\sim {m_E}\right)$ to $2\pi T$ approximately. We present the scale dependence of $m_{\rm E}^2$ before and after applying the PMC in Fig.\ref{fig:mE}. It shows that the PMC prediction on $m_{\rm E}^2$ is independent to the choice of $\mu_r$, whose value under conventional scale-setting shows a non-negligible scale dependence~\footnote{A similar discussion on the EFT parameter has been done by using the prototype of PMC, i.e. the Brodsky-Lepage-Mackenzie (BLM) scale-setting~\cite{Brodsky:1982gc}, and our corresponding PMC scales are consistent with the BLM predictions~\cite{Braaten:1995jr, Braaten:1995ju}.}. We present the scale dependence of $m_{\rm E}^2$ by using three typical scales $\pi T/2$, $\pi T$ and $2\pi T$ in Table~\ref{tab:mE}. Under conventional scale-setting, $m_{\rm E}^2$ varies by $[-7\%, +2\%]$ when $\mu_r\in[\pi T/2, 2\pi T]$.

\begin{figure}[htb]
\centering
\includegraphics[width=0.5\textwidth]{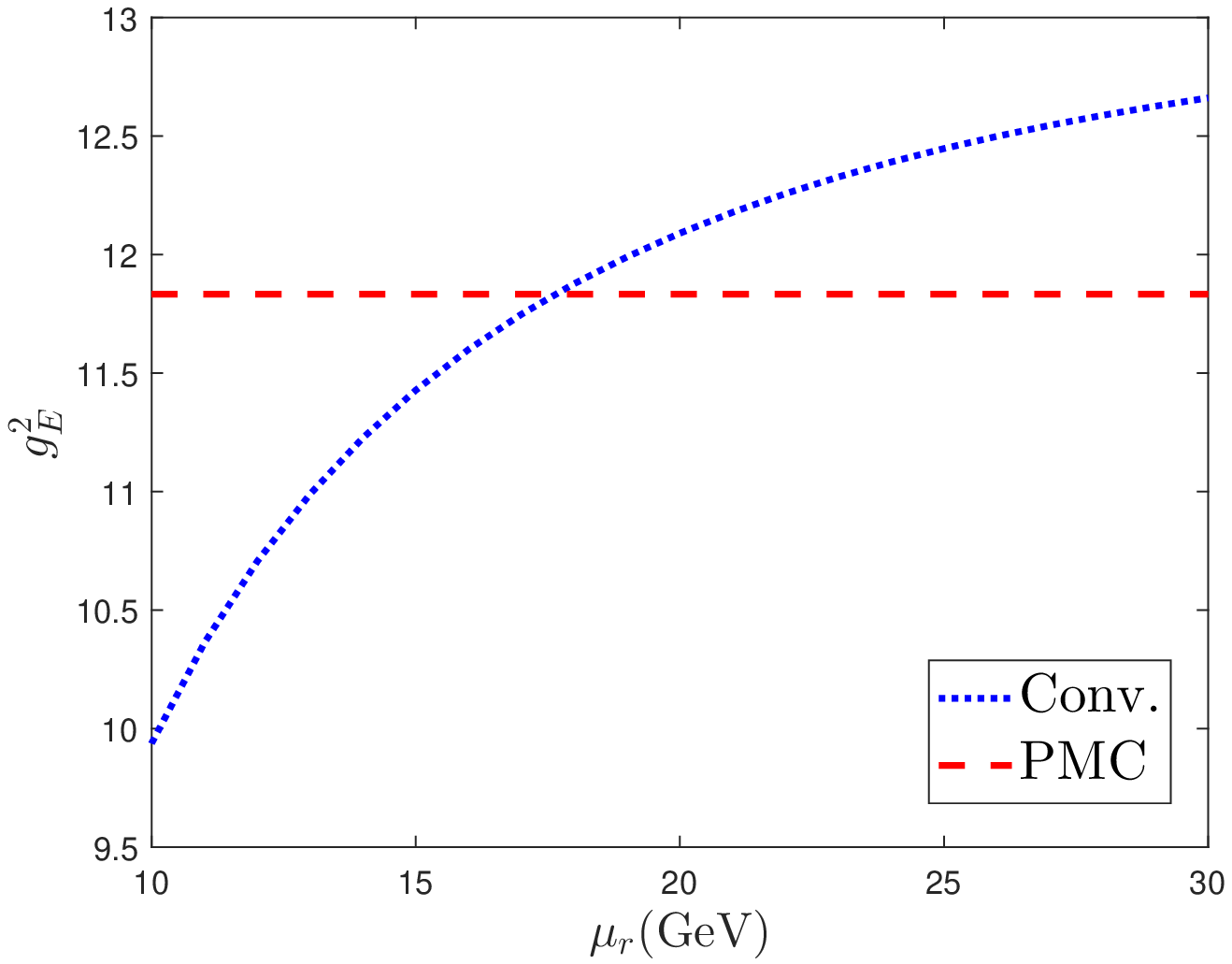}
\caption{The EFT parameter $g_{\rm E}^2$ versus the renormalization scale ($\mu_r$) under conventional and PMC scale-settings, respectively. $T=10$ GeV.}
\label{fig:gE}
\end{figure}

\begin{table}[htb]
\begin{center}
\begin{tabular}{c ccccc}
\hline
$g_{\rm E}^2$~ & ~${\mu _r}$ ~ & ~\rm{LO}~ & ~\rm{NLO}~ & ~Total~  \\
\hline
& ${\pi T/4}$ & $21.53$ & $-12.94$ & $8.59$  \\
~\rm{Conv.}~ & ${\pi T/2}$ & $18.12$ & $-6.57$ & $11.55$  \\
& ${\pi T}$ & $15.68$ & $-2.98$ & $12.70$  \\
\hline
~\rm{PMC}~ & $\left[ {\pi T/4,\pi T} \right]$ & $17.65$ & $-5.82$ & $11.83$  \\
\hline
\end{tabular}
\caption{The results of $g_{\rm E}^2$ under conventional and PMC scale-settings, respectively. Three typical scales, $\mu_r=\pi T/4$, $\pi T/2$, and $\pi T$, are adopted. $T=10$ GeV.}
\label{tab:gE}
\end{center}
\end{table}

As mentioned in Sec.II, for a ${\cal O}(g_s^5)$-order prediction on the free energy density, we only need a LO $g_E^2$. However by using the LO $g_E^2$ alone, we cannot determine its renormalization scale. To achieve a more accurate prediction on $g_E^2$ itself, we adopt the known NLO-terms~\cite{Kajantie:2002wa} to set the scale for $m_E^2$. The scale dependence of $g_{\rm E}^2$ up to NLO level before and after applying the PMC scale-setting is presented in Fig.\ref{fig:gE}, which shows the scale dependence can be eliminated by applying the PMC. Numerical results for $g_{\rm E}^2$ under three typical scales $\pi T/4$, $\pi T/2$ and $\pi T$ are presented in Table~\ref{tab:gE}. It shows that $g_{\rm E}^2$ under conventional scale-setting varies by $[-26\%, +10\%]$ for $\mu_r\in[\pi T/4,\pi T]$.

\begin{table}[tb]
\begin{center}
\begin{tabular}{c cccccc}
\hline
${F_M}/{F_{\rm ideal}}$~ & ~$\mu_r$ ~ & ~ LO~ & ~NLO~ & ~NNLO~ & ~Total~ \\
\hline
       & ${\pi T/4}$ & $0.07$ & $-0.15$ & $-0.16$ & $-0.24$ \\
       & ${\pi T/2}$ & $0.10$ & $-0.15$ & $-0.13$ & $-0.18$ \\
$\rm{Conv.}$ & ${\pi T}$ & $0.12$ & $-0.14$ & $-0.10$ & $-0.12$ \\
       & ${2\pi T}$ & $0.12$ & $-0.13$ & $-0.08$ & $-0.09$ \\
\hline
PMC-I  & $\left[ {1/4\pi T,2\pi T} \right]$ & $0.11$ & $-0.14$ & $-0.10$ & $-0.13$ \\
PMC-II  & $\left[ {1/4\pi T,2\pi T} \right]$ & $0.11$ & $-0.16$ & $ -0.12$ & $-0.17 $ \\
\hline
\end{tabular}
\caption{The ratio ${F_M}/{F_{\rm ideal}}$ under conventional and PMC scale-settings, respectively. Four typical scales, ${\mu_r} = \pi T/4$, $\pi T/2$, $\pi T$ and $2\pi T$, are adopted. The LO $g_{\rm E}^2$ with $Q_1^g=0.93\pi T$ (PMC-I) or $Q_1^g=0.56\pi T$ (PMC-II) is adopted for a ${\cal O}(g_s^5)$-order prediction. $T=10$ GeV.}
\label{tab:FMgQm}
\end{center}
\end{table}

As a summary, by substituting the EFT parameters $m_{\rm E}^2$ and $g_E^2$ into Eq.(\ref{FMlam}), we obtain the PMC prediction for the ratio ${F_M}/{F_{\rm ideal}}$, which are presented in Table~\ref{tab:FMgQm}. Summing the $F_E$ and ${F_M}$ together,  by taking $Q_1^g \equiv Q_1^m=0.93\pi T$ to calculate the LO $g_{\rm E}^2$,  we obtain
\begin{equation}
\frac{F}{F_{\rm ideal}}\bigg|_{T=10GeV} = 0.866_{ - 0.002}^{ + 0.003}.
\end{equation}
where the uncertainty is for $\Delta{\alpha _s}({1.5{\rm GeV}}) =\left({}_{-0.008}^{ + 0.012}\right)$. If taking $Q_1^m=0.56\pi T$ determined from the known NLO $g_E^2$-term to calculate the LO $g_{\rm E}^2$, we obtain
\begin{equation}
\frac{F}{F_{\rm ideal}}\bigg|_{T=10GeV} = 0.827_{ - 0.003}^{ + 0.004}.
\end{equation}

\section{Summary}
\label{sec:summary}

In the paper, we have studied the properties of the free energy density up to $g_s^5$-order at the high temperature $T$ by applying the PMC within the EFT framework. The PMC provides a systematic method to set the renormalization scale of the high-energy process, whose predictions are free of renormalization scale dependence even for low-order predictions. As shown by Tables \ref{tab:FE} and \ref{tab:FMgQm}, our predictions on the free energy density up to $g_s^5$-order confirm this observation.

\begin{figure}[htb]
\centering
\includegraphics[width=0.5\textwidth]{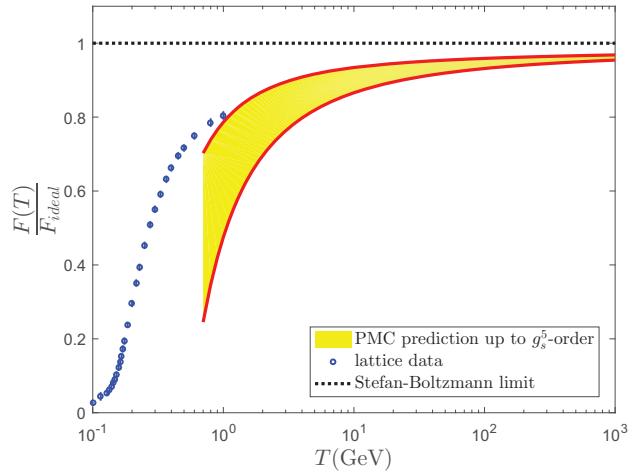}
\caption{The PMC prediction of the free energy density up to $g_s^5$-order versus the temperature $T$ with free quark numbers in QGP $n_f=3$. The upper edge of the band corresponds to $\Lambda_E=m_E$ and the lower edge of the band corresponds to $\Lambda_E=2\pi T$. The lattice data with pion mass $m_\pi=160$ GeV~\cite{Borsanyi:2010cj} and the Stefan-Boltzmann limit of the ideal gas are presented as a comparison.}
\label{fig:FTZ}
\end{figure}

It is noted that the determination of the factorization scale is a completely separate issue from the renormalization scale setting problem, since it is presented even for a conformal theory with $\beta=0$. With the help of Eqs.(\ref{eq:Fpmc}, \ref{eq:Fpmc2}), we present a prediction on the factorization scale dependence on the ratio $F/F_{\rm ideal}$ up to $g_s^5$-order as a function of $T$ in Fig.\ref{fig:FTZ}. The factorization scale uncertainty is discussed by taking the range, $ {g_s T}\sim m_E <{\Lambda_E} < {2\pi T}$~\cite{Cvetic:2004xq}. The dashed line indicates the Stefan-Boltzmann limit of the ideal gas. The lattice data for the case $n_f=2+1$~\cite{Borsanyi:2010cj} is adopted for a comparison. The upper edge of the band corresponds to $\Lambda_E=m_E$ and the lower edge of the band corresponds to $\Lambda_E=2\pi T$. Fig.\ref{fig:FTZ} shows that when $\Lambda_E=m_E$, the free energy density agrees with the lattice data even for low temperature $T$ around $1$ GeV, indicating a smaller factorization scale is more preferable.

\hspace{1cm}

{\bf Acknowledgements}: This work was supported in part by the Natural Science Foundation of China under Grant No.11625520.

\end{document}